# Identifying and correcting invalid citations due to DOI errors in Crossref data

Alessia Cioffi, alessia.cioffi@studio.unibo.it, https://orcid.org/0000-0002-9812-4065
Digital Humanities and Digital Knowledge, Department of Classical Philology and Italian Studies, University of Bologna, Bologna, Italy

Sara Coppini, sara.coppini@studio.unibo.it, https://orcid.org/0000-0002-6279-3830
Digital Humanities and Digital Knowledge, Department of Classical Philology and Italian Studies, University of Bologna, Bologna, Italy

Arcangelo Massari, arcangelo.massari@unibo.it, https://orcid.org/0000-0002-8420-0696
Research Centre for Open Scholarly Metadata, Department of Classical Philology and Italian Studies, University of Bologna, Bologna, Italy

Arianna Moretti, arianna.moretti2@studio.unibo.it, https://orcid.org/0000-0001-5486-7070
Digital Humanities and Digital Knowledge, Department of Classical Philology and Italian Studies, University of Bologna, Bologna, Italy

Silvio Peroni, silvio.peroni@unibo.it, https://orcid.org/0000-0003-0530-4305
Research Centre for Open Scholarly Metadata, Department of Classical Philology and Italian Studies, University of Bologna, Bologna, Italy
Digital Humanities Advanced Research Centre (/DH.arc), Department of Classical Philology and Italian Studies, University of Bologna, Bologna, Italy

Cristian Santini, Cristian.Santini@fiz-Karlsruhe.de, https://orcid.org/0000-0001-7363-6737
FIZ Karlsruhe – Leibniz Institute for Information Infrastructure, Karlsruhe, Germany
Karlsruhe Institute of Technology, Institute AIFB, Karlsruhe, Germany

Nooshin Shahidzadeh Asadi, nooshin.shahidzadehasadi@uantwerpen.be, https://orcid.org/0000-0003-4114-074X
Antwerp Centre for Digital Humanities and Literary Criticism, University of Antwerp, Antwerp, Belgium

## Abstract

This work aims to identify classes of DOI mistakes by analysing the open bibliographic metadata available in Crossref, highlighting which publishers were responsible for such mistakes and how many of these incorrect DOIs could be corrected through automatic processes. By using a list of invalid cited DOIs gathered by OpenCitations while processing the OpenCitations Index of Crossref open DOI-to-DOI citations (COCI) in the past two years, we retrieved the citations in the January 2021 Crossref dump to such invalid DOIs. We processed these citations by keeping track of their validity and the publishers responsible for uploading the related citation data in Crossref. Finally, we identified patterns of factual errors in the invalid DOIs and the regular expressions needed to catch and correct them.

The outcomes of this research show that only a few publishers were responsible for and/or affected by the majority of invalid citations. We extended the taxonomy of DOI name errors proposed in past studies and defined more elaborated regular expressions that can clean a higher number of mistakes in invalid DOIs than prior approaches. The data gathered in our study can enable investigating possible reasons for DOI mistakes from a qualitative point of view, helping publishers identify the problems underlying their production of invalid citation data. Also, the DOI cleaning mechanism we present could be integrated into the existing process (e.g. in COCI) to add citations by automatically correcting a wrong DOI.

This study was run strictly following Open Science principles, and, as such, our research outcomes are fully reproducible.

# 1. Introduction

Citations are one of the fundamental aspects of scientific research since they enable acknowledging the work of others, showing its provenance, and affirming the reliability of the claims and data used in a study. Hence, in the past few years, several initiatives adopted faster, more transparent, available, and reliable practices in research. For instance, the San Francisco Declaration on Research Assessment (DORA, https://sfdora.org/) and Plan-S (https://www.coalition-s.org/) have demanded open access to scholarly articles and their metadata, including articles' reference lists, following the practice started with the Initiative for Open Citations (I4OC, https://i4oc.org/). Recently, I4OC has successfully convinced academic publishers to make their reference lists available in Crossref (Hendricks et al., 2020) (https://crossref.org) as public domain data. Several projects and organisations have started to reuse Crossref citation data once made available through their REST APIs (https://api.crossref.org). Among them, OpenCitations (Peroni & Shotton, 2020) built COCI, the OpenCitations Index Of Crossref Open DOI-to-DOI Citations (COCI, https://opencitations.net/index/coci) (Heibi et al., 2019b).

However, not all the DOI-to-DOI citations that can be gathered from Crossref are transferred into COCI. Since the publishers themselves provide all metadata in Crossref and Crossref does not double-check such data, the ingestion process developed for creating COCI validates each DOI involved in a citation using the API of the DOI.org service (https://doi.org/api/handles/). In case a DOI involved in a citation is incorrect (i.e. it does not exist), such citation is not added in COCI to prevent including erroneous information in the dataset.

The identification of the sources and classes of such DOI mistakes have been addressed in prior studies. Several articles describe in detail common mistakes made by publishers concerning DOIs. For instance, Valderrama-Zurián et al. (2015) and Franceschini et al. (2016) focus on the "errors/horrors" present in the Scopus database, while Zhu et al. (2019) describe the errors inherent in the Web of Science (WoS) database. Instead, Gorraiz et al. (2016) and Franceschini et al. (2015) analyse this problematic aspect in both the WoS and the Scopus databases. Finally, Boudry & Chartron (2017) deal with the mistakes recorded in the PubMed database.

Along the same lines, Buchanan (2006) proposed a taxonomy of the main DOI errors, distinguishing between author errors and database mapping errors. The former can be inaccuracies caused by authors while creating the list of articles cited for their publications; the latter can be attributed to a data entry error, which raises failures in creating an electronic link between a cited article and the corresponding citing article. Another kind of taxonomy was recently proposed by Xu et al. (2019): they studied the data in Web of Science and identified three distinct classes of errors, i.e. errors in the DOI prefix (e.g. the presence of the HTTP or HTTPS scheme in the DOI definition), suffix errors (e.g. the presence of queries to the proxy server, hashes, and delimiters), and other-type errors (e.g. double underscores, double periods, XML tags, spaces, and forward slashes). In addition, the authors proposed a regular expression-based methodology to correct these three errors, which will be reused and refined in this work. A similar study, using a similar collection of data, was also presented by Zhu et al. (2019).

In this article, we want to accompany the outcomes mentioned above with new insights and perspectives on this issue concerning DOI mistakes, using citation data and bibliographic metadata available in open datasets to make the outcomes of our research fully reproducible. In particular, we wanted to answer the following research questions (RQ1 and RQ2 from now on):

1. Which publishers were responsible (due to the incorrect DOI metadata they sent to Crossref) for the missing citations in COCI?
2. What were the classes of errors that characterised invalid DOIs, and how much of these DOIs could be corrected through automatic processes?

To answer these two questions, we have devised an open methodology (introduced in Section 2) compliant with the FAIR principles (Wilkinson et al., 2016): it uses open platforms and tools to define and share the whole research workflow, from processing the input data to producing final outcomes. The results obtained using this methodology are presented in Section 3, followed (in Section 4) by a discussion of such results, highlighting their significance and limits. Finally, Section 5 concludes the work by summarising the answers to RQ1-RQ2 and outlining some future works.

# 2. Material and methods

This section will present the processes, tools, and software used to gather and analyse the data to answer RQ1 and RQ2. After introducing the initial input data (Section 2.1), we will describe the two workflows devised to answer RQ1 (Section 2.2) and RQ2 (Section 2.3), respectively. In order to double-check the material created to implement each workflow, we split ourselves into two groups: G1 and G2. Each group was responsible for the creation and implementation of one of the two workflows and related data. The material produced by each group was carefully reviewed by the other group to guarantee high-quality standards of the produced material.

## 2.1. Input data

A list of invalid cited DOIs was gathered by OpenCitations while processing COCI in the past two years. Starting from the citation data available in Crossref as of January 2021 (Crossref, 2021), a list of all the citations in Crossref to the invalid DOIs identified during the COCI processing activities was created and made available online (Peroni, 2021). The data consists of a CSV file, where each row is a pair of *valid citing DOI – invalid cited DOI* (see sample in Table 1). The file contains a total of 1,223,295 DOI-to-DOI invalid citations.

| Valid citing DOI | Invalid cited DOI |
|---|---|
| 10.14778/1920841.1920954 | 10.5555/646836.708343 |
| 10.5406/ethnomusicology.59.2.0202 | 10.2307/20184517 |
| 10.1177/1179546820918903 | 10.3748/wjg.v10.i5.707. |

*Table 1: Sample of the CSV containing Crossref citations to invalid DOI (Peroni, 2021).*

## 2.2. Identifying publishers responsible for invalid DOIs

G1 processed the input CSV file and created a new dataset, keeping track of each citation's validity and the publisher responsible for uploading the related citation data in Crossref. We queried the DOI.org API (https://doi.org/api/handles/) with the DOI of the cited entity involved in a citation to check whether the citation was currently valid (i.e. the invalid DOI gathered via the COCI process has changed its status to valid in the meantime). We assumed that the citing DOI was correct since Crossref provided it.

Conversely, we used the Crossref API (https://api.crossref.org) to identify publishers of both the citing and the cited entities involved in each citation by employing their DOIs' prefixes, i.e. the part of the DOI starting with "10." and ending before the first "/" (e.g. "10.1016"). In case a publisher could not be identified in Crossref we labelled it as "unidentified" in the main results. In cases where this failure of recognition was deemed to be the result of publishers not having used Crossref as their DOI registration agency (implied by doi.org recognizing a DOI as valid but Crossref rejecting it), we used API services of two other agencies, DataCite and mEDRA, and URL recognition for one other agency, CNKI, in order to recognize such publishers. The result of this analysis was saved separately so as not to pollute the main results that were directly garnered from Crossref.

All the material produced by G1 is available at https://github.com/open-sci/2020-2021/blob/master/docs/TheLeftovers20/material.md. The Data Management Plan (Cioffi et al., 2021a), the description of the protocol to gather and analyse the data (Coppini et al., 2021b), the software implementing the protocol (Coppini et al., 2021a), and the final data obtained by running the workflow (Cioffi et al., 2021b) are all available online to enable the reproducibility of the study.

## 2.3 Classifying and correcting DOIs classes of errors

G2 developed an approach to classify invalid DOIs and clean the related invalid citation data. The cited invalid DOIs provided in the input data can be either (a) temporarily invalid or (b) factually invalid. In (a), the cited DOIs were invalid when the COCI ingestion workflow was executed, and they may have become valid before our analysis. The analysis started months after the publication of the COCI releases from which invalid DOIs have been taken, and we found that some DOIs classified as invalid in the data were then returned as valid when querying the DOI API. In (b), the DOIs needed to be corrected due to, for instance, the presence of an additional character.

| Id | Example of invalid DOI | Example of correct DOI | Error | Regular expression | In (Xu et al. 2019) |
|---|---|---|---|---|---|
| 1 | 10.1016/J.AMEPRE.2015.07.017. | 10.1016/J.AMEPRE.2015.07.017 | Suffix | (.*?)(?:[\.\,\|<|&|\(|;]+)$ | modified |
| 2 | 10.1186/1735-2746-10-21,HTTP://WWW.IJEHSE.COM/CONTENT/10/1/21 | 10.1186/1735-2746-10-21 | Suffix | (.*?)(?:[\.\(|,|;]*HTTP:\/\/.*?)$ | modified |
| 3 | 10.1016/J.JLUMIN.2004.10.018.HTTP://DX.DOI.ORG/10.1016/J.JLUMIN.2004.10.018 | 10.1016/J.JLUMIN.2004.10.018 | Prefix | (.*?)(?:\.)?(?:HTTP:\/\/DX\.D[0|O]I\.[0|O]RG\/)(.*)$ | modified |
| 4 | 10.1093/BIOINFORMATICS/BTV421.HTTPS://DOI.ORG/10.1101/GR.186072.114 | 10.1093/BIOINFORMATICS/BTV421 | Prefix | (.*?)(?:\.)?(?:HTTPS:\/\/D[0|O]I\\.[0|O]RG\/)(.*)$ | modified |
| 5 | 10.1016/J.TIBS.2006.12.007....32,63(2006) | 10.1016/J.TIBS.2006.12.007 | Suffix | (.*?)(?:\.{5}.*?)$ | no |
| 6 | 10.1021/BI3013565(2012) | 10.1021/BI3013565 | Suffix | (.*?)(?: \(\d{4}\)?)$ | yes |
| 7 | 10.1287/ORSC.2016.1092>ACCESSED27 | 10.1287/ORSC.2016.1092 | Suffix | (.*?)(?:[>\)](LAST)?ACCESSED\d+)$ | no |
| 8 | 10.1111/J.1536-7150.2006.00482.X/FULL>ACCESSED4 | 10.1111/J.1536-7150.2006.00482.X | Suffix | (.*?)(?:\/(META\|ABSTRACT\|FULL\|EPDF\|PDF\|SUMMARY)([>\|\)](LAST)?ACCESSED\d+)?)$ | no |
| 9 | 10.1007/3-540-35074-8_16#PAGE-1 | 10.1007/3-540-35074-8_16 | Suffix | (.*?)(?: #.*?)$ | no |
| 10 | 10.1371/JOURNAL.PONE.0112567.PMID:25405489 | 10.1371/JOURNAL.PONE.0112567 | Suffix | (.*?)(?:[\.\(|,|;]?PMID:\d+.*?)$ | modified |
| 11 | 10.1063/1.1148310?CRAWLER=TRUE | 10.1063/1.1148310 | Suffix | (.*?)(?: \?.*?=.*?)$ | no |
| 12 | 10.1177/0004865814524218ANJ.SAGEPUB.COM | 10.1177/0004865814524218 | Suffix | (.*?)(?:[\.\(|,|;]?[A-Z]\*\.?SAGEPUB.*?)$ | no |
| 13 | 10.1073/PNAS.1104391108[DOI] | 10.1073/PNAS.1104391108 | Suffix | (.*?)(?: \[DOI\].*?)$ | no |
| 14 | 10.1073/PNAS.1319051111/-/DCSUPPLEMENTAL | 10.1073/PNAS.1104391108 | Suffix | (.*?)(?:\/-\/DCSUPPLEMENTAL)$ | yes |
| 15 | 10.1890/15-0075.1/SUPPINFO | 10.1890/15-0075.1 | Suffix | (.*?)(?:\/SUPPINF[0|O](\.)?)$ | modified |
| 16 | 10.1101/GR.229202.ARTICLEPUBLISHEDONLINEBEFOREMARCH2002 | 10.1101/GR.229202 | Suffix | (.*?)(?:[\.\(|,|;]?ARTICLEPUBLISHEDONLINE.*?\d{4})$ | modified |
| 17 | 10.1016/J.JPROT.2014.03.043(EPUBAHEADOFPRINT) | 10.1016/J.JPROT.2014.03.043 | Suffix | (.*?)(?:[\(|\[]EPUBAHEADOFPRINT[\)\|]])$ | yes |
| 18 | 10.1016/j.chom.2007.09.014,PMCID:PMC2184509 | 10.1016/j.chom.2007.09.014 | Suffix | (.*?)(?:[\.\|\(|,|;]?PMCID:PMC\\d+.*?)$ | yes |
| 19 | 10.1186/1471-2407-13-87<br/> | 10.1186/1471-2407-13-87 | Other | (?:<.*?/>) | yes |
| 20 | 10.3390/v4061011\\ | 10.3390/v4061011 | Other | (?:\\)\\ | yes |
| 21 | 10.1007/978-3-319-04765-2_2 | 10.1007/978-3-319-04765-2_2 | Other | (?:_)_ | yes |
| 22 | 10.1111/j.1540-4560..2011.01712.x | 10.1111/j.1540-4560.2011.01712.x | Other | (?:\.)\. | yes |
| 23 | 10.1037/0022-<xml_add>e</xml_add>3514.52.3.511 | 10.1037/0022-3514.52.3.511 | Other | (?:<.*?>.*?</.*?>) | yes |

*Table 2: List of recurrent factual errors in invalid DOIs. In the second column, the wrong part of the considered DOIs is highlighted in red. The examples are accompanied by the respective classes of error and regular expressions (case insensitive) used to catch them. The red part of the regular expression identifies the portion of the invalid DOI that will be removed. We also specified if that regular expression was already listed in (Xu et al., 2019).*

For further classification of the invalid DOIs, we reused the categorisation presented by Xu et al. (2019), which split all the DOIs in (b) as follows:

- DOIs with *prefix-type* errors, i.e. when they have additional characters *before* the specific pattern sequence "10.XXXX/XXX..." (e.g. "http://dx.doi.org/10.1016/j.aca.2006.07.086");
- DOIs with *suffix-type* errors, i.e. when they have additional characters *after* the specific sequence "10.XXXX/XXX…" (e.g. "10.1186/1735-2746-10-21,http://www.ijehse.com/content/10/1/21").
- DOIs with *other type* errors, i.e. they do not comply with the previous categories and contain wrongly added characters, such as double slashes, double underscores, or XML tags (e.g. "10.1007/978-3-319-04765-2__2").

Patterns of factual errors were manually checked by processing the input data. Table 2 includes all the error types of the identified type (b) invalid DOIs, accompanied by examples and the regular expressions needed to catch them. Our regular expressions were implemented by using the Regular Expression library available in Python, which follows a syntax similar to Perl's. Complete documentation is available on the Python website (https://docs.python.org/3/library/re.html). Moreover, Appendix 1 provides an overview of the rules used.

We slightly modified some of the regular expressions used in (Xu et al., 2019) to find other patterns in the input data. In particular, a pattern was added to capture the new public DOI proxy server (i.e. "https://doi.org"), in addition to the still supported previous syntax (i.e. "http://dx.doi.org"), which is no longer preferred. Furthermore, in the regular expression for cleaning prefix errors, the character "O" and the number "0" (zero) were both counted as a match since they are often confused, as shown in (Zhu et al., 2019). Nine new patterns were devised to clean errors in suffixes. As to the preexisting regular expressions for suffix errors, some minor changes were made, for instance, in the occurrence and type of delimiters between the DOI and the erroneous part of the string at the end. Finally, regarding other-type errors, substitutions were made to remove the wrong characters or to replace the repetitions, such as double underscores, with a single character.

We extended Xu et al. (2019)'s cleaning method, which is based on regular expressions for removing DOI errors in Web of Science. In particular, we enriched it with additional patterns that occurred in our input data (as shown in Table 2), implementing the additional checks described in steps 1 and 3 of the following list:

1. We processed the input data and checked, using the DOI.org API, whether any of the cited DOIs were currently valid before starting the cleaning procedure. All the DOIs retrieved as valid identifiers were removed from the following steps of our analysis;
2. The remaining invalid DOIs were cleaned using the regular expressions introduced in Table 2. In particular, we created one regular expression to clean prefix-type errors, one for cleaning suffix-type errors, and another for other types of error.  Following the example by Xu et al. (2019), the regular expressions of the same type, introduced in Table 2, were combined as alternatives into a non-capturing group – e.g. for the prefix-type error, the final regular expression used was `(.*?)(?:\.)?(?:HTTP:\/\/DX\.D[0|O]I\.[0|O]RG\/)(.*))|(.*?)(?:\`

`.)?(?:HTTPS:\/\/D[0|O]I\.[0|O]RG\/)(.*).` These groups of regular expressions were executed in the order specified above (i.e. prefix-type first, then suffix-type and finally other types). If a regular expression returned two cleaned strings – e.g. as shown in the third and fourth row of Table 2 – we kept the longest one. This process allowed us to produce a new cleaned version of the invalid DOIs;

3. The validity of each cleaned DOI was rechecked with the DOI.org API. We stored the new current status of the cleaned DOIs either as *valid* or *invalid* depending on the result gathered from the API.

Finally, we checked the correctness of the cleaned DOIs to identify possible mistakes introduced by the correction – e.g. when a DOI does not refer to any of the bibliographic resources cited by a particular citing article. We randomly selected 10 citations per pattern from the corrected citations obtained by the process described above. Examining a fixed quantity of citations per regular expression and not an amount proportional to the number of matches removed the bias given by the population under consideration. Then, we manually checked each of the cleaned DOIs by retrieving the text of the original citing articles from the DOIs included in the input data. Thus, we recorded whether the related citing articles, if accessible, included bibliographic references to the cleaned DOIs. In view of this result, we measured the accuracy of the cleaning process.

All the material produced by G2 is available at https://github.com/open-sci/2020-2021/blob/master/docs/Grasshoppers/material.md. The Data Management Plan (Boente et al. 2021b), the protocol to gather and analyse the data (Boente et al., 2021c), the software implementing the protocol (Massari et al., 2021), the final data (Boente et al., 2021a), and the evaluation data (Massari, 2022) are all available online to enable the reproducibility of the study.

# 3. Results

One of the primary purposes of our work (RQ1) was to identify the publishers responsible for uploading invalid citations (i.e. bibliographic references of citing articles having specified wrong DOIs) and, along the same line, which publishers are potentially affected by such mistakes. Figure 1[1] shows the twenty publishers having the highest number of invalid outgoing citations according to the data in (Peroni, 2021), and also shows the portion of outgoing citations that, in the meantime, became valid (according to our queries to the DOI.org API). The publisher with the most invalid outgoing citations was Ovid Technologies, with more than 370,000 invalid outgoing citations, followed by Springer, the Association for Computing Machinery, Informa UK Limited, and Wiley. On the other hand, the outgoing citations that became correct after being discarded from COCI were distributed among a few publishers in the past two years. In particular, 71% of invalid outgoing citations in (Peroni, 2021) from Elsevier citing articles became valid, as well as 82% of invalid outgoing citations from Cambridge University Press citing articles.

---

[1] All the figures introduced in this section are also available online at https://open-sci.github.io/2020-2021-the-leftovers-20-code and https://open-sci.github.io/2020-2021-grasshoppers-code.

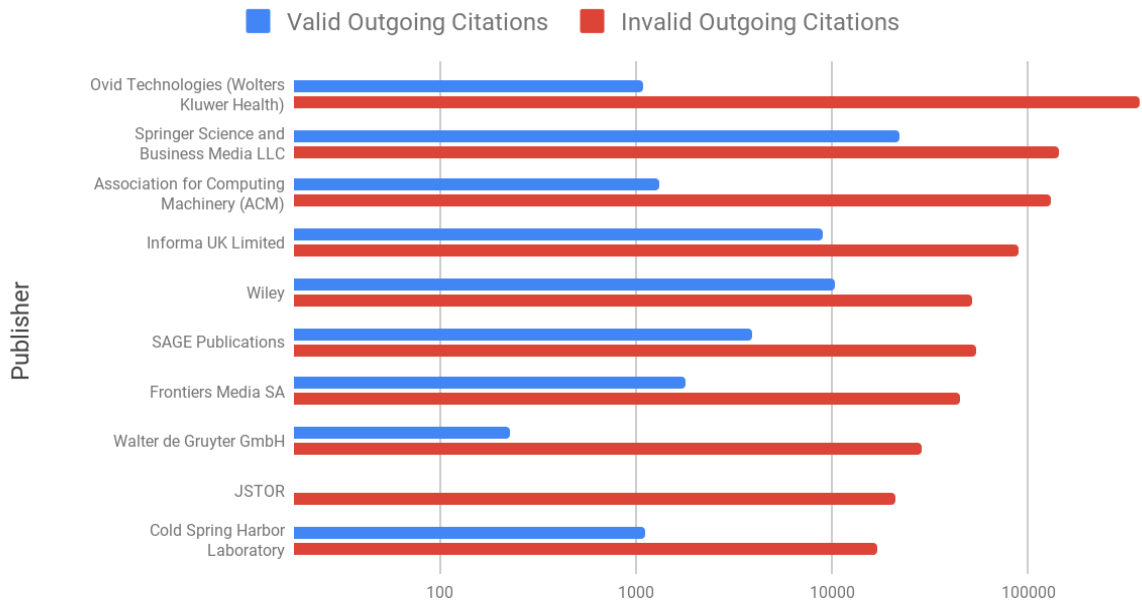

*Figure 1: Top ten publishers responsible for the highest number of invalid outgoing citations in (Peroni, 2021), listed in descending order. **This data was** collected before the cleaning process.*

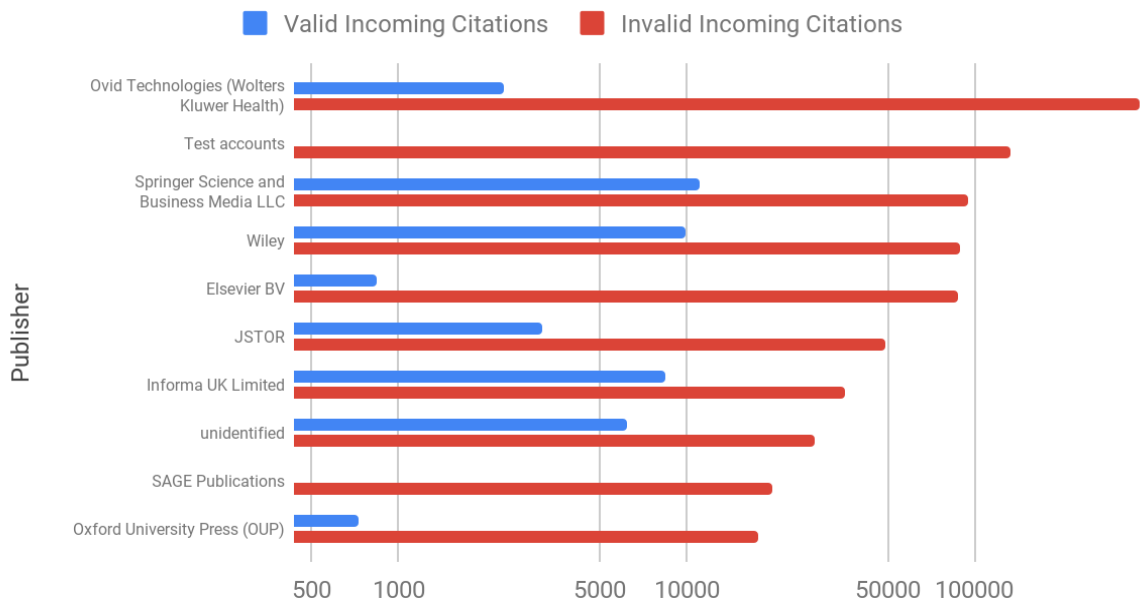

*Figure 2: Top ten publishers affected by the highest number of invalid incoming citations in (Peroni, 2021), listed in descending order. **This data was** collected before the cleaning process.*

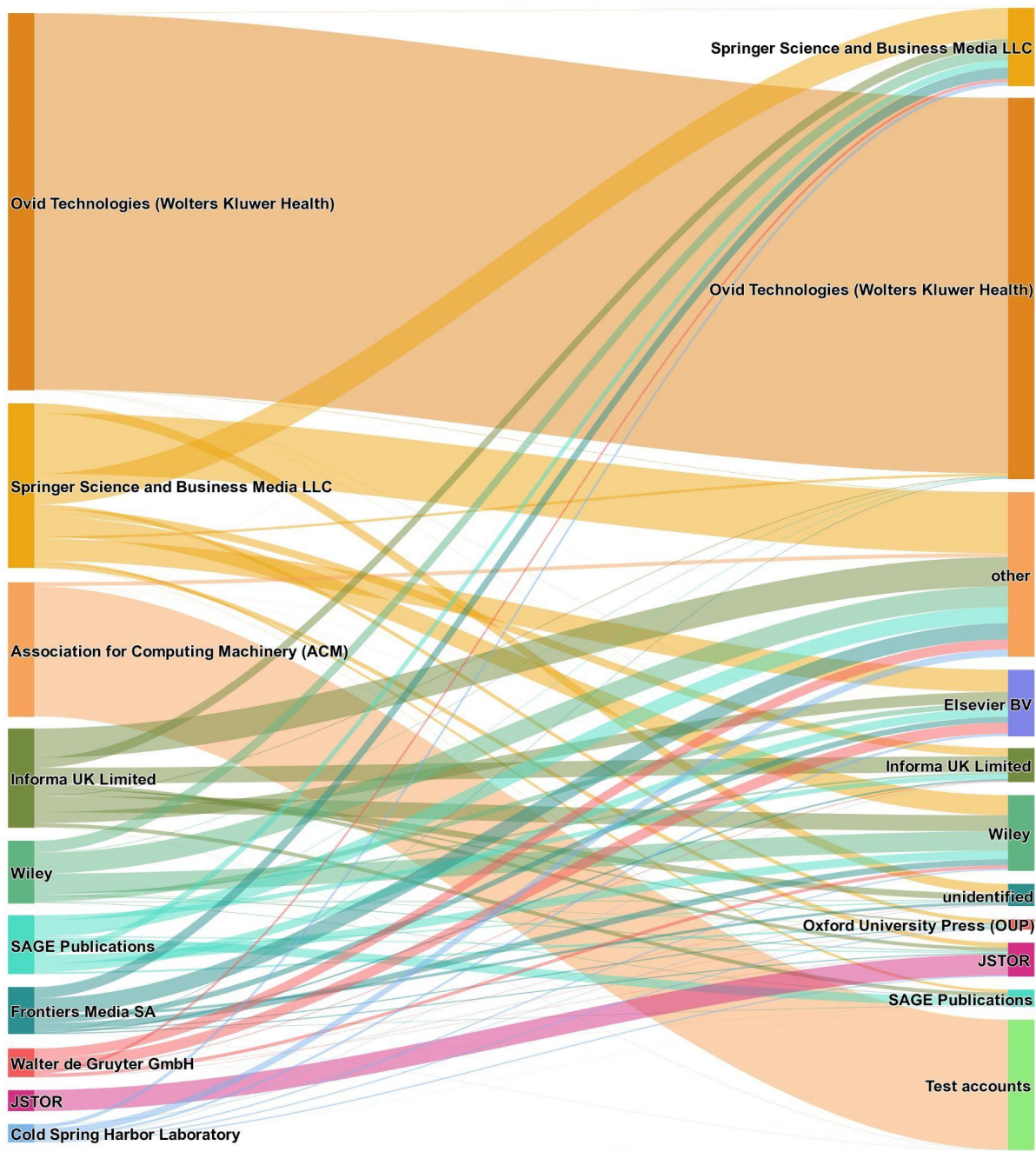

*Figure 3: Sankey diagram showing the top ten publishers responsible for invalid outgoing citations (left) as in (Peroni, 2021) and the related publishers pointed by such invalid citations (right). The label "other" is used as a residual category.*

Figure 2 shows the publishers that were affected the most by invalid citations, distinguishing citations that become valid after the dataset's creation from those that remained invalid. Again, Ovid Technologies was the publisher with the highest number of invalid incoming citations (about 380,000 invalid incoming citations). Other relevant cited publishers are Test accounts (a special account created, in Crossref, for testing purposes), Springer, Wiley, and Elsevier. Once more, the number of invalid incoming citations in (Peroni, 2021) which became valid after querying the DOI.org API involved a few selected publishers. Among these, Walter de Gruyter recorded the highest number of previously invalid, and now valid, incoming citations (77%). Both Figure 1 and Figure 2 display data that was collected before the correction phase.

Several publishers are present in both Figure 1 and Figure 2. Notably, Ovid Technologies was by far the first in both cases – which made us speculate about a possible case of publisher self-citations, although such citations were invalid. To further investigate this hypothesis, we reorganised the data shown in Figures 1 and 2, considering the publishers cited by the first ten publishers responsible for invalid outgoing citations, as shown in the Sankey diagram in Figure 3. Indeed, it seems that almost all the invalid outgoing citations in Ovid Technologies' and JSTOR's citing articles pointed to Ovid Technologies' and JSTOR's cited resources, respectively. Also, the invalid outgoing citations of the citing articles published by the Association for Computing Machinery showed an interesting pattern: most of such invalid citations pointed to the Test account.

| Id | Occurrences |
|---|---|
| 1 | 114,948 |
| 2 | 21,816 |
| 3 | 21,695 |
| 8 | 2,890 |
| 6 | 859 |
| 20 | 818 |
| 22 | 588 |
| 5 | 464 |
| 13 | 82 |
| 10 | 79 |
| 12 | 60 |
| 7 | 57 |
| 4 | 47 |
| 19 | 34 |
| 21 | 33 |
| 9 | 26 |
| 11 | 15 |
| 14 | 13 |
| 16 | 8 |
| 17 | 4 |
| 23 | 1 |
| 15 | 0 |
| 18 | 0 |

*Table 3: The identifiers of the DOI error types reported in Table 2 and their occurrences, considering the invalid citations DOIs in (Peroni, 2021), which were previously invalid but became valid after being cleaned.*

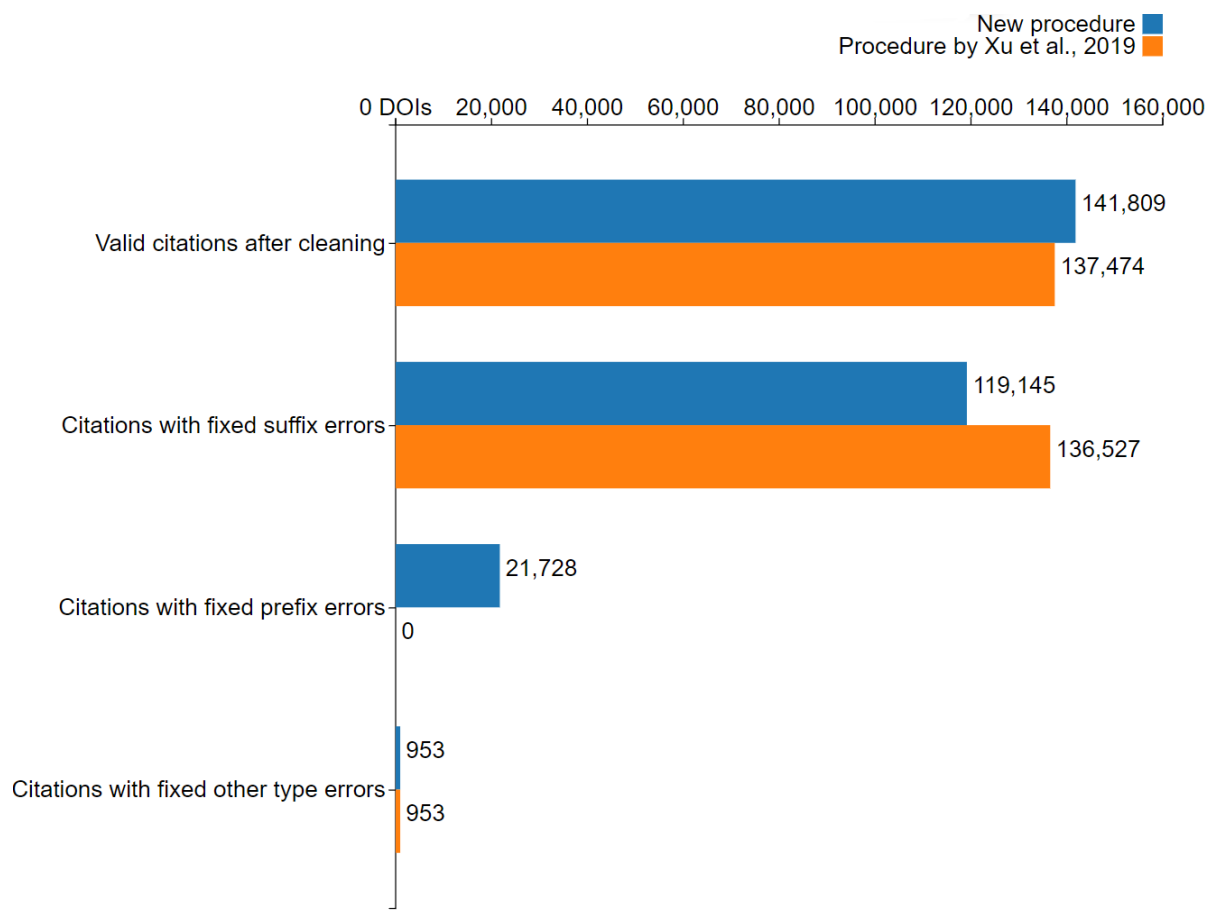

*Figure 4: Bar chart showing the results obtained with both the reference procedure described by Xu et al. (orange) and the extended procedure described in this research article (blue).*

Once the publishers responsible or affected by invalid citations were identified, we focused on recognising the classes of errors that characterised invalid DOIs in (Peroni, 2021). Moreover, we quantified the number of citations that may become valid by automatically correcting part of these invalid DOIs (RQ2). Table 3 shows how many occurrences out of the patterns in Table 2 were identified in the cited DOIs of the citations in (Peroni, 2021). Only citations that proved valid after being cleaned and that are currently invalid were considered in the count.

We developed some code to implement the DOI rewriting shown in Table 2 and introduced in Section 2.3. Then, we measured the number of citations we fixed (i.e., that became valid) by having such correction code executed. As shown in Figure 4, 141,809 citations were cleaned by the methods presented in this study, corresponding to 11,6% of the whole dataset. In particular, 119,145 citations became valid by fixing a suffix error of the cited entities' DOI, constituting 84% of the citations that became valid after applying our rewriting method. This number exceeds the 57,196 citations that were temporarily invalid during the COCI ingestion workflow and that were marked as valid before applying any cleaning method. The prefix errors we automatically corrected via our computational methods comprised 15,3% of all the citations that became valid after cleaning, while other-type errors could only be found rarely in the input dataset.

We compared our method for cleaning invalid DOIs to the one introduced by Xu et al. (2019), and the results are introduced in Figure 4. It is worth mentioning that it was impossible to precisely reproduce Xu et al.'s procedure as the source code used in their study was not made available. Consequently, the comparison was made using their set of regular expressions in our pipeline and checking how many of the DOIs cleaned with Xu et al.'s regular expressions were returned as valid by the DOI.org API. It emerged that 137,474 citations were corrected by their methodology, which is 4,335 less than our extended procedure.

Finally, as anticipated in Section 2.3, we manually checked 10 random fixed citations for each of the 23 regular expressions to measure the quality of the cleaning approach proposed. As a result, we obtained 193 citations to be manually checked. We did not get 230 citations because the patterns 15, 16, 17, 18 and 23, all derived from the article by Xu et al., have less than ten matches in the input dataset. The cleaning of 191 of these 193 citations proved correct, one proved incorrect, while the article identified by the other citing DOI was paywalled and not accessible to us because it was published in a journal not included in our institution's contracts.

# 4. Discussions

Having analysed the results, we noticed that only a few publishers were responsible for providing invalid DOIs in their articles' references (Figure 1). Although many of the publishers mentioned above were also largely affected by invalid incoming citations, the large-scale publishers, as noted in (Ruediger Wischenbart Content and Consulting, 2019), were the ones receiving the highest number of invalid incoming citations (Figure 2). For instance, while Elsevier was not responsible for several invalid outgoing citations (Figure 1), it was largely impacted by a huge amount of invalid incoming citations (Figure 2) by several other publishers (Figure 3). This phenomenon could be due to the massive number of articles Elsevier publishes in a year, which most likely resulted in having a more significant probability of being cited.

Looking at the cited entities' DOIs in the set of invalid citations, we found that a large number of such DOIs had a prefix that was ascribable to a publisher not included in Crossref. In Figure 3, we used the umbrella term "unidentified" to group all invalid incoming citations pointing to these unidentified publishers. Specifically, we gathered 4,618 unidentified publishers in our dataset that do not use Crossref as a DOI provider for their publications. This collection of prefixes was then analysed for membership in other DOI registration agencies, such as mEDRA, DataCite, and CNKI, and many were recognized to be, in fact, valid publishers. Information regarding these publishers and their names was saved in the output file.

Along the same lines, we identified a peculiar publisher in Crossref named "Test accounts", having the prefix "10.5555". The name of this hypothetical publisher, the lack of information about it on Crossref, and the fact that it received only invalid incoming citations suggests that this publisher was merely used for testing purposes and, as such, it does not exist. However, it is worth mentioning that it ranked second in Figure 2 with more than 17,000 invalid incoming citations, mainly from ACM publications, as shown in Figure 3.

The Sankey diagram in Figure 3 shows that the number of invalid incoming citations was very high only for a few publishers, which were also the ones responsible for the majority of the invalid outgoing citations. These numbers suggest that such publishers cited themselves using invalid – or at least, not yet validated – cited DOIs. The most prominent publisher among them is Ovid Technologies, which exceeded the others by a large margin, despite being smaller, in terms of the number of publications per year, than Elsevier and Springer. Even though the reasons behind this practice are out of the scope of our study, they are definitely worth further research.

As shown in Figure 4, we were able to assess as valid 16% of the invalid citation data included in (Peroni, 2021). This percentage includes many DOIs that became valid over time without being cleaned (5%) and those that became valid after our cleaning methodology was applied. The largest part of the cleaned DOIs originally contained suffix-type errors, which were the most prominent type of errors among the cited DOIs in our input dataset. However, it is worth noting that the regular expression for suffix-type errors was broader than those for the other types of errors since it included nine different rules, as shown in Table 2.

Furthermore, as shown in Table 3, the number of matches regarding suffixes and other type errors is higher than the amount of DOIs cleaned with those regular expressions. In fact, a DOI can contain more than one type of error simultaneously, and our procedure solves all of them in sequence. For example, considering the invalid DOI "10.1016/j.sbspro.2014.01.467<br/>http://www.sciencedirect.com", both the URL (rule 2 in Table 3) and the "<br/>" tag (rule 19 in Table 3) were removed.

While assessing our cleaning mechanism's quality on 193 invalid randomly chosen citations, we noticed some peculiarities with the two citations we could not prove correct. A closer investigation showed that their still-incorrect state is not to be imputed to the cleaning methodology developed. In particular, in one of these invalid citations, the DOI "10.1007/978-3-319-90698-0_26" (Krebs, 2018) should be included in the references of the article having DOI "10.17660/actahortic.2020.1288.20" (Wang et al., 2020), which, unfortunately, is behind a paywall. Our institution (i.e. the University of Bologna) was not subscribed to the journal containing that article (i.e. *ISHS Acta Horticulturae*) at the time of writing. Conversely, the other invalid citation included the article having DOI "10.1007/s10479-011-0841-3." as cited by the article having DOI "10.1101/539833" (Domanskyi et al., 2019). Checking the citing article, we realised the cited work associated with the invalid DOI is not the one our regex corrected by removing the trailing dot (García-Alonso et al., 2014), but another one that does not have a DOI (van der Maaten & Hinton, 2008). In other words, there was no way to fix the invalid DOI, as there is no correct DOI. Interestingly, van der Maaten and Hinton's work is cited 118 times in the input dataset, always with the same non-existent DOI. A manual verification highlighted that 115 of these erroneous citations were traceable to bioRxiv, while Gut, MCP, and JBC presented one incorrect citation each. Since the article published on MCP was initially filed as a pre-print on bioRxiv (Meng et al., 2019), the number of wrong pointers originating from bioRxiv is 116. Among them, two works (García-Timermans et al., 2019; Li et al., 2019) cite the DOI fixed with our regex but associate it with the article by van der Maaten and Hinton, which is not the right one. It is plausible that there was a preprocessing error in biorXiv, and that the wrong information contained therein was also the source for other publishers.

We also compared our cleaning methodology with the one introduced by Xu et al. (2019). As shown in Figure 8, our process was able to correct a larger number of DOIs overall: 141,809 citations proved valid, not considering the citations already valid, i.e. validated over time and not because of correction or cleaning processes. On the other hand, 137,474 citations were found to be valid with the method by Xu et al. (2019). This improvement was primarily due to the addition of the eighth pattern in Table 2, which alone corrected 2,889 more citations. However, the method by Xu et al. (2019) allowed the cleaning of 17,395 more suffix-type errors than our approach. Nevertheless, it did not clean any prefix errors in the given dataset. This result can be explained by taking a closer look at the regular expressions and Xu et al.'s definition of prefix-type and suffix-type errors. In our study, the regular expressions for prefix errors were broader since we considered any additional strings before the DOI as prefix errors, even if they were not at the beginning of the string. Conversely, Xu et al.'s method considered prefix errors only those at the beginning of the string, hence the absence of matches in the given dataset. For instance, in the DOI "10.1016/J.JLUMIN.2004.10.018.HTTP://DX.DOI.ORG/10.1016/J.JLUMIN.2004.10.018", our method considered the presence of the DOI proxy server (i.e. "HTTP://DX.DOI.ORG/") as a prefix-error, because it appeared before the DOI. Presumably, such errors were fixed as suffix errors in Xu et al.'s method, leading to more matches for this kind of error.

Indeed, the different choices in our and Xu et al.'s studies show that defining classes of errors is a task open for interpretation, and seemingly small decisions can lead to dissimilar results. A manual investigation of the input dataset is required to define regular expressions, implying that such different error types depend on the input dataset and the scholars extracting and analysing information. Therefore, the regular expressions used in our work can be used to clean only a part of the possible existing errors that can be retrieved in the literature.

# 5. Conclusions

Our article provided a methodology for discovering the publishers responsible for submitting incorrect references' DOIs to Crossref (RQ1), which resulted in missing citations in OpenCitations' COCI. This work also identified the classes of errors that characterised such invalid DOIs and corrected them through automatic processes (RQ2). We noticed that only a few publishers are responsible for and/or affected by the majority of invalid citations and that only a tiny part of the invalid citations in (Peroni, 2021) had become valid before our analysis. We also extended the taxonomy of DOI name errors proposed by Xu et al. (2019) and defined more elaborated regular expressions that can clean a higher amount of mistakes in invalid DOIs. The cleaning of the incorrect DOIs of the analysed invalid citations was particularly effective in the presence of suffix-type errors, but such corrections were strongly dependent on the input data used. In principle, the coverage of the cleaning rules we introduced can be extended if additional sources of invalid DOIs are provided.

The data generated by our work and its outcomes have multiple implications that can be considered in future studies. Firstly, we could investigate possible trends behind DOI mistakes from a qualitative perspective, helping publishers identify the problems underlying their production of invalid citation data. Secondly, the cleaning mechanism we introduced could be integrated into the COCI ingestion process to add missed citations due to a mistake in the cited DOI that could be corrected automatically. Finally, we could propose some strategies and mechanisms to provide the publishers with the cleaned version of formerly incorrect DOIs, to prevent the insertion of incorrect citation data in repositories such as Crossref.

# Acknowledgements

We want to thank the anonymous reviewers of this article and (in alphabetic order) Sebastian Barzaghi, Ricarda Boente, Marilena Daquino, Silvia Di Pietro, Ivan Heibi, Bruno Sartini, and Deniz Tural, who provided insightful feedback and contributed to some of the aspects of the present study. This study was based on the outcomes of the analysis held during the Open Science 2020/2021 course (https://www.unibo.it/en/teaching/course-unit-catalogue/course-unit/2020/443753) of the Second Cycle International Degree in Digital Humanities and Digital Knowledge (https://corsi.unibo.it/2cycle/DigitalHumanitiesKnowledge) of the University of Bologna. This work has been partially funded by the European Union's Horizon 2020 research and innovation program under grant agreement No 101017452 (OpenAIRE-Nexus).

# Appendix 1

## Regular Expressions

Regular Expressions are a series of characters defining a sequence of strings. This sequence can be specified either by defining the exact strings in order or by using a series of special and non-special characters defining compositional operations between characters.
We listed below the subset of components and rules of the Regular Expressions syntax used in our software, providing specific examples.

## Special Symbols

In regular expressions, most of the characters match themselves. For example, the regular expression `House` will match only the exact string `"House"`. However, this is not true for all characters. The special characters included in our cleaning method are:

- `.` – the character class for matching any character except a newline;
- `$` – the anchor for matching the end of the string. More in detail, it matches the position of the sequence of characters specified. Thus, `$` is not aimed at matching the characters' sequence, but rather at checking the condition that a sequence of characters ends.

## Quantifiers

### Exact Quantifier (`{n}`)

The quantifier specifies how many repetitions are required of the sequence of characters it is attached to. The quantifier `{n}` exactly matches *n* repetitions of a given sequence; `{n,m}` matches from *n* to *m* repetitions of the previous sequence; and `{n, }` matches at least *n* repetitions of the sequence. For example, in the regular expression `(.*?)(?:[\.|\(|,|;]?ARTICLEPUBLISHEDONLINE.*?\d{4})$`, the exact quantifier `{4}` after the escaped "d" character is aimed at matching the four digits `\d` representing a year.

### Star (`*`)

Matches 0 or more repetitions of the characters sequence it is attached to. The sequence can be either absent or present, and if present it can be repeated any number of times.

### Plus (`+`)

Matches at least one occurrence of the sequence of characters it is attached to. It means that at least one occurrence of the sequence must be present, and it can be repeated any number of times.

Optional (`?`)

Matches at most one occurrence of the sequence it refers to. The presence of the sequence of characters is optional and, if it is present, it must not be repeated.

Greedy and Lazy Quantifiers

The quantifiers are greedy. They match as many characters as possible unless they have to apply the rest of the regular expression. Suppose you have the regular expression `<.+>`; this regular expression will match the whole string `"<p>Hello World</p>"`. By putting a question mark after the quantifier, it will make it lazy; therefore, the regular expression `<.+?>` will match only `"<p>"` inside the string `"<p>Hello World</p>"`.

# Character Sets (`[xyz]`)

Character sets match only one character out of a set of characters. For example, the character set `[aeiou]` will match only one among the vowels `"a"`, `"e"`, `"i"`, `"o"`, and `"u"`. Character sets were used in our regular expressions, combined with the end of string operator `$`, to match a set of unwanted characters at the end of a DOI. For example, the regular expression `(.*?)(?:[\.,<&\(;]+)$` will match an invalid DOI having a dot or a comma at the end.

# Alternation (`[x|y]`)

The alternation between two characters or series of characters matches either the preceding or the element following the vertical line symbol and thus has the same function of the boolean operator OR. The characters or combinations of characters are checked in order of appearance, i.e., the one on the left of the alternation symbol first, and the one on the right after. For example, in the regular expression `(.*?)(?:\.)?(?:HTTPS:\/\/D[0|O]I\.[0|O]RG\/)(.*)$`, the alternation between `0` and `O` (i.e., `[0|O]`), addresses the possibility that the URL `"HTTPS://DOI.ORG"` was mistyped as `"HTTPS://D0I.ORG"`, `"HTTPS://DOI.0RG"`, or `"HTTPS://DOI.ORG"`.

# Escaped Characters (`\x`)

In the regular expressions language, some characters (i.e., `+*?^$\.[]{}()|/`) have a symbolic function, and thus are by default interpreted for their functional meaning. However, they may be included in a regular expression to match exact characters. In this case, a backslash character (`\`) placed before the character signifies the intention to escape their functional role and consider them as literal characters.

In addition to the set of characters listed above, it is also possible to escape the literal value of certain alphabetic characters to represent classes of characters. For example, the letter `w` exactly matches a lowercase `"w"` character. In contrast, `\w` matches a word character. Similarly, `d` exactly matches a lowercase `"d"` character, while `\d` matches a digit character; `s` exactly matches a lowercase `"s"` character, while `\s` matches a spacing character. Therefore, in the regular expression

`(.*?)(?:[\.|\(|,|;]?ARTICLEPUBLISHEDONLINE.*?\d{4})$`, `\d` stands for *digit* and not for the sequence of a backslash and a lowercase `"d"` character.

## Capturing Groups

A capturing group gathers multiple elements together and matches them as a single group. Groups can be nested, and thus a capturing group can result from the sequence of capturing groups contained in the parentheses.

## Non-Capturing Group

A non-capturing group gathers multiple elements together without creating a capturing group. This option is preferred to the capturing group when the element following the `?:` symbol is searched, but there is no need to capture the value for further recalls. Indeed, non-capturing groups were used in most of our regular expressions to match and remove unwanted characters either at the beginning or the end of a DOI. For example, the incorrect DOI `"10.1371/JOURNAL.PONE.0112567.PMID:25405489"` was cleaned by using the regular expression `(.*?)(?:[\.|\(|,|;]?PMID:\d+.*?)$`, where the non-capturing group `(?:[\.|\(|,|;]?PMID:\d+.*?)` serves the purpose of removing the unwanted error.